\documentclass[preprint,showpacs,preprintnumbers,amsmath,amssymb,superscriptaddress,nofootinbib]{revtex4-1}

\usepackage{graphicx}
\usepackage{dcolumn}
\usepackage{bm}
\usepackage{tabularx}
\usepackage{color}

\newcolumntype{L}{>{\raggedright\arraybackslash}X}
\newcolumntype{C}{>{\centering\arraybackslash}X}
\newcolumntype{R}{>{\raggedleft\arraybackslash}X}

\begin{document}

\preprint{APS/123-QED}

\title{Hindered proton collectivity in $^{28}_{16}$S$^{}_{12}$: Possible
magic number at $Z=16$}

\author{Y.~Togano}
\thanks{Present address: EMMI, GSI, D-64291 Darmstadt, Germany}
\affiliation{RIKEN Nishina Center, Saitama 351-0198, Japan}
\affiliation{Department of Physics, Rikkyo University, Tokyo 171-8501, Japan}%
\author{Y.~Yamada}
\affiliation{Department of Physics, Rikkyo University, Tokyo 171-8501, Japan}%
\author{N.~Iwasa}
\affiliation{Department of Physics, Tohoku University, Miyagi 980-8578, Japan}%
\author{K.~Yamada}
\affiliation{RIKEN Nishina Center, Saitama 351-0198, Japan}%
\author{T.~Motobayashi}
\affiliation{RIKEN Nishina Center, Saitama 351-0198, Japan}%
\author{N.~Aoi}
\affiliation{RIKEN Nishina Center, Saitama 351-0198, Japan}%
\author{H.~Baba}
\affiliation{RIKEN Nishina Center, Saitama 351-0198, Japan}%
\author{S.~Bishop}
\affiliation{RIKEN Nishina Center, Saitama 351-0198, Japan}%
\author{X.~Cai}
\affiliation{Shanghai Institute of Applied Physics, Chinese Academy of
Science, Shanghai 201800, China}%
\author{P. Doornenbal}
\affiliation{RIKEN Nishina Center, Saitama 351-0198, Japan}%
\author{D.~Fang}
\affiliation{Shanghai Institute of Applied Physics, Chinese Academy of
Science, Shanghai 201800, China}%
\author{T.~Furukawa}
\affiliation{RIKEN Nishina Center, Saitama 351-0198, Japan}%
\author{K.~Ieki}
\affiliation{Department of Physics, Rikkyo University, Tokyo 171-8501, Japan}%
\author{T.~Kawabata}
\affiliation{Center for Nuclear Study, University of Tokyo, Saitama
351-0198, Japan}%
\author{S.~Kanno}
\affiliation{RIKEN Nishina Center, Saitama 351-0198, Japan}%
\author{N.~Kobayashi}
\affiliation{Department of Physics, Tokyo Institute of Technology,
Tokyo 152-8551, Japan}
\author{Y.~Kondo}
\affiliation{RIKEN Nishina Center, Saitama 351-0198,
Japan}%
\author{T.~Kuboki}
\affiliation{Department of Physics, Saitama University, Saitama 338-8570,
Japan}%
\author{N.~Kume}
\affiliation{Department of Physics, Tohoku University, Miyagi 980-8578, Japan}
\author{K.~Kurita}
\affiliation{Department of Physics, Rikkyo University, Tokyo 171-8501, Japan}%
\author{M.~Kurokawa}
\affiliation{RIKEN Nishina Center, Saitama 351-0198, Japan}%
\author{Y.~G.~Ma}
\affiliation{Shanghai Institute of Applied Physics, Chinese Academy of
Science, Shanghai 201800, China}%
\author{Y.~Matsuo}
\affiliation{RIKEN Nishina Center, Saitama 351-0198, Japan}%
\author{H.~Murakami}
\affiliation{RIKEN Nishina Center, Saitama 351-0198, Japan}%
\author{M.~Matsushita}
\affiliation{Department of Physics, Rikkyo University, Tokyo 171-8501, Japan}%
\author{T.~Nakamura}
\affiliation{Department of Physics, Tokyo Institute of Technology,
Tokyo 152-8551, Japan}
\author{K.~Okada}
\affiliation{Department of Physics, Rikkyo University, Tokyo 171-8501, Japan}%
\author{S.~Ota}
\affiliation{Center for Nuclear Study, University of Tokyo, Saitama 
351-0198, Japan}
\author{Y.~Satou}
\affiliation{Department of Physics, Tokyo Institute of Technology,
Tokyo 152-8551, Japan}
\author{S.~Shimoura}
\affiliation{Center for Nuclear Study, University of Tokyo, 
Saitama 351-0198, Japan}
\author{R.~Shioda}
\affiliation{Department of Physics, Rikkyo University, Tokyo 171-8501, Japan}%
\author{K.~N.~Tanaka}
\affiliation{Department of Physics, Tokyo Institute of Technology,
Tokyo 152-8551, Japan}
\author{S.~Takeuchi}
\affiliation{RIKEN Nishina Center, Saitama 351-0198, Japan}%
\author{W.~Tian}
\affiliation{Shanghai Institute of Applied Physics, Chinese Academy of
Science, Shanghai 201800, China}%
\author{H.~Wang}
\affiliation{Shanghai Institute of Applied Physics, Chinese Academy of
Science, Shanghai 201800, China}%
\author{J.~Wang}
\affiliation{Institute of Modern Physics, Chinese Academy of Science, 
Lanzhou 730000, China}%
\author{K.~Yoneda}
\affiliation{RIKEN Nishina Center, Saitama 351-0198, Japan}%


\date{\today}

\begin{abstract}
 The reduced transition probability $B$(E2;$0^+_{gs} \rightarrow 2^+_1$)
 for $^{28}$S was obtained experimentally using Coulomb excitation
 at 53~MeV/nucleon. 
 The resultant $B$(E2) value 181(31)~e$^2$fm$^4$ is smaller than the
 expectation based on empirical $B$(E2) systematics.
 The double ratio $|M_n/M_p|/(N/Z)$ of the $0^+_{gs} \rightarrow 2^+_1$ 
 transition in $^{28}$S was determined to be 1.9(2) by
 evaluating the $M_n$ value from the known $B$(E2) value of the mirror
 nucleus $^{28}$Mg,
 showing the hindrance of proton collectivity relative to 
 that of neutrons.
 These results indicate the emergence of the magic number $Z=16$ in the 
 $|T_z|=2$ nucleus $^{28}$S.
\end{abstract}

\pacs{23.20.Js, 25.60.-t, 25.70.De}
\maketitle

%
Magic numbers characterize the shell structure of fermionic quantum
system such as atoms, metallic clusters \cite{Cluster} and nuclei 
\cite{Nuclei}.
A unique feature of the nuclear system is the fact that it comprises
two types of fermions, the protons and neutrons, 
and hence the magic numbers appear both for protons and neutrons.
Most of the recent studies regarding the magic numbers are for
neutron-rich nuclei.
Disappearance of the conventional magic numbers of $N$=8, 20 and 28 
\cite{Iwasaki, Warburton, Bastin}
or the appearance of 
the new magic number $N=16$ \cite{Ozawa, Kanungo, Hoffman} has been
shown.
They are associated with nuclear collectivity, which is enhanced, for
instance, in the neutron-rich $N=20$ nucleus $^{32}$Mg caused by
disappearance of the magic number \cite{Detraz, Motobayashi}.

The new neutron magic number $N=16$ has been confirmed experimentally for 
$^{27}$Na ($|T_z|=5/2$) and more neutron-rich isotones 
\cite{Ozawa, Hoffman, Kanungo, Gibelin, Cooper}.
Its appearance can be theoretically interpreted as 
a result of a large gap
between the neutron $d_{3/2}$ and $s_{1/2}$ orbitals caused by
the low binding energy \cite{Ozawa} and/or the spin-isospin dependent 
part of the residual nucleon-nucleon interaction \cite{Otsuka}.
In analogy to the magic number $N=16$, the proton magic number $Z=16$
must also exist in proton-rich nuclei.
However, it has not been identified experimentally in the
proton-rich sulfur isotopes.
The present Letter reports on a study of the magic number $Z=16$ at the
most proton-rich even-even isotope $^{28}$S with $|T_z|=2$ through 
a measurement of the reduced transition probability 
$B$(E2;$0^+_{gs} \rightarrow 2^+_1$).

The $B$(E2) value is directly related to the 
amount of quadrupole collectivity of protons.
The relative contribution of the proton- and neutron-collectivities can be 
evaluated using the ratio of the neutron transition matrix element 
to the proton one (the $M_n/M_p$ ratio) for $0^+_{gs} \rightarrow 2^+_1$
transitions \cite{Bernstein, Bernstein3}.
$M_p$ is related to $B$(E2) by
$e^2 M_p^2 = B({\rm E2};0^+_{gs} \rightarrow 2^+_1)$.
The $M_n$ value can be deduced from the $M_p$ value in the mirror
nucleus, where the numbers of protons and neutrons are interchanged.
If collective motions of protons and neutrons have the same amplitudes,
the double ratio $|M_n/M_p|/(N/Z)$ is, therefore, expected to be unity.
Deviation from $|M_n/M_p|/(N/Z) = 1$ corresponds to a proton/neutron
dominant excitation and should indicate a difference in the 
motions of protons and neutrons.
Such a difference appears typically for
the singly-magic nuclei \cite{Bernstein, Kennedy}.
For proton singly-magic nuclei, the proton collectivity is
hindered by the magicity, leading to $|M_n/M_p|/(N/Z) > 1$.
For example, the singly-magic nucleus $^{20}$O has a large double ratio
of $1.7 \sim 2.2$ for the $0^+_{gs} \rightarrow 2^+_1$ transition 
\cite{Jewell, Kahn, Iwasa}.

We used Coulomb excitation at an intermediate energy to extract the 
$B$(E2;$0^+_{gs} \rightarrow 2^+_1$) value of the proton-rich nucleus 
$^{28}$S.
Intermediate-energy Coulomb excitation is a powerful tool to obtain 
$B$(E2) with relatively low intensity beams because a thick target 
is available \cite{Motobayashi, Glasmacher}.
The double ratio $|M_n/M_p|/(N/Z)$ of the $0^+_{gs} \rightarrow 2^+_1$
transition 
is obtained by combining the $B$(E2) values of $^{28}$S and the
mirror nucleus $^{28}$Mg.

The experiment was performed using the RIBF (Radioactive Isotope Beam Factory)
accelerator complex operated by RIKEN Nishina Center and Center for
Nuclear Study, University of Tokyo.
A $^{28}$S beam was produced via projectile fragmentation of a
115-MeV/nucleon $^{36}$Ar beam from the $K=540$~MeV RIKEN Ring Cyclotron 
incident on a 531~mg/cm$^2$ thick Be target. 
The secondary beam was obtained by the RIKEN
Projectile-fragment separator (RIPS) \cite{RIP1992KU00} using an aluminum
energy degrader with a thickness of 221~mg/cm$^2$ and a wedge angle of
1.46~mrad placed at the first dispersive focus.
The momentum acceptance was set to be $\pm$ 1\%.
A RF deflector system \cite{NSR2004YA29} was placed at 
the second focal plane of RIPS
to purify the $^{28}$S in the beam with intense contaminants 
(mostly of $^{27}$P, $^{26}$Si and $^{24}$Mg) 
that could not be
removed only by the energy loss in the degrader.
Particle identification for the secondary beam was performed
event-by-event by measuring time of flight (TOF), energy loss 
($\Delta E$), and the magnetic rigidity of each nucleus.
TOF was measured by using a RF signal from the cyclotron and a 0.1~mm
thick plastic scintillator located 103~cm upstream of the third focal
plane.
$\Delta E$ was obtained by a 0.1~mm thick silicon detector placed 117~cm
upstream of the third focal plane.
The average $^{28}$S beam intensity was 120 s$^{-1}$, which corresponded
to approximately 1.9\% of the total intensity of the secondary beam.
The secondary target was a 348~mg/cm$^2$-thick lead sheet which was 
set at the third focal plane.
The average beam energy at the center of the lead target was
53~MeV/nucleon. 
Three sets of PPACs \cite{PPAC} were placed 155.6~cm, 125.6~cm, and
66.2~cm upstream of the secondary target, respectively,
to obtain the beam trajectory on the secondary target.

An array of 160 NaI(Tl) scintillator crystals, DALI2 \cite{DALI2}, 
was placed
around the target to measure de-excitation $\gamma$ rays from 
ejectiles.
The measured full energy peak efficiency was 30 \% at 0.662~MeV, in
agreement with a Monte-Carlo simulation made by the {\small GEANT4} code, 
and the energy resolution was 9.5 \% (FWHM).
The full-energy-peak efficiency for 1.5~MeV $\gamma$ rays emitted from 
the ejectile with the velocity of $0.32c$ was evaluated to be 16\% 
by the Monte-Carlo simulation.

The scattering angle, energy loss ($\Delta E$), and total energy ($E$)
of the ejectiles from
the lead target were obtained by a detector telescope located 62~cm downstream
of the target.
It consisted of four layers of silicon detectors arranged in
a $5 \times 5$ matrix without 4 detectors at the corners for the first
two layers, and a $3 \times 3$ matrix for the third and fourth
layers.
The silicon detectors in the four layers had an effective area of
$50 \times 50$~mm$^2$ and a thickness of 500, 500, 325, and
500~$\mu$m, respectively.
The detectors in the first and second layers had 5-mm-wide strip
electrodes on one side to determine the hit position of the ejectiles.
The $\Delta E$-$E$ method was employed to identify $^{28}$S.
The mass number resolution for sulfur isotopes was 0.35 
(1$\sigma$).
The angle of the ejectile was obtained from the hit position on the
telescope and the beam angle and position on the target measured by the
PPACs.
The scattering angle resolution was 0.82 degree.

\begin{figure}[t]
 \includegraphics[clip,keepaspectratio,width=0.44\textwidth]{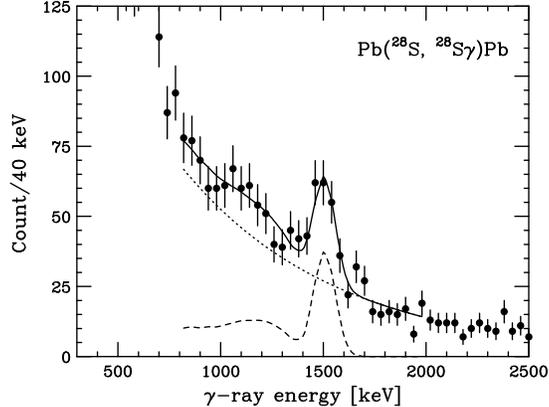}
 \caption{\label{fig:gamma} Doppler-shift corrected $\gamma$-ray 
 energy-spectrum in the Pb($^{28}$S,$^{28}$S $\gamma$)Pb reaction.
 The fit by the response function (dashed curve) and the exponential
 background (dotted curve) is shown by the solid curve. 
}
\end{figure}

The Doppler-shift corrected $\gamma$-ray energy-spectrum measured in
coincidence with inelastically scattered $^{28}$S is shown in
Fig. \ref{fig:gamma}. 
A peak is clearly seen at 1.5~MeV. 
The spectrum was fitted by a detector response 
obtained by the Monte-Carlo simulation and an exponential background.
The peak energy was obtained to be 1.497(11)~MeV, which was
consistent with the previous measurement, 1.512(8)~MeV, 
by the two neutron removal reaction on $^{30}$S \cite{Yoneda}.
This peak has been assigned to the transition from the $2^+_1$ state to the
$0^+$ ground state \cite{Yoneda}.
In extracting the inelastic cross section, transitions feeding the
$2^{+}_1$ state were not accounted for, because the proton
separation energy of 2.46(3)~MeV is relatively low and no higher
excited states were seen in the present spectrum and the
two-neutron removal reaction on $^{30}$S \cite{Yoneda}.
This was supported by the location of the second excited state in the mirror
nucleus $^{28}$Mg of 3.86~MeV.

\begin{figure}[t]
 \includegraphics[clip,keepaspectratio,width=.44\textwidth]{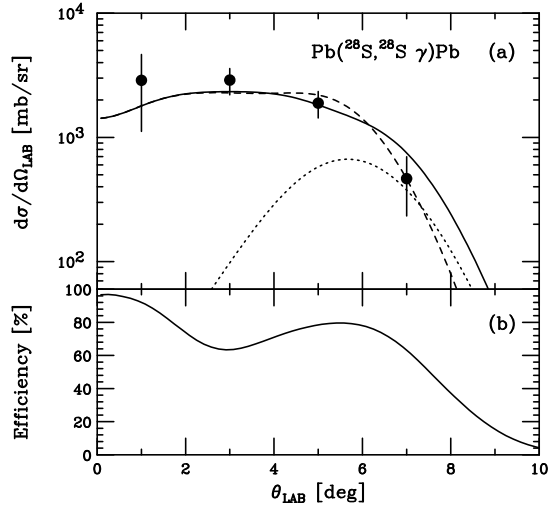}
 \caption{\label{fig:dsdomega} (a) Angular distribution for the 
Pb($^{28}$S,$^{28}$S $\gamma$)Pb reaction exciting the 1.5~MeV state in
$^{28}$S. The solid curve represents the best fit with {\small ECIS} 
 calculation assuming $\Delta L=2$. 
 The dashed and dotted curves show the Coulomb and nuclear contributions, 
 respectively.
 (b) Detection efficiency calculated by the Monte-Carlo simulation.
}
\end{figure}

The angular distribution of the scattered $^{28}$S excited to its
1.5~MeV state is shown in Fig. \ref{fig:dsdomega}(a).
Figure \ref{fig:dsdomega}(b) shows the angle-dependence of the detection
efficiency for scattered $^{28}$S obtained by a Monte-Carlo simulation.
It took into account the spacial and angular distributions of the
$^{28}$S beam, the size of the silicon detectors, and effect of multiple
scattering in the target.
The cross section integrated up to 8 degree was obtained to be 99(16)~mb
by taking into account the angle-dependent detection efficiency.
The error was nearly all attributed to the statistical uncertainties,
while the systematic errors of the $\gamma$-ray detection
efficiency and the angle-dependence of the detection efficiency were
also included (3\%).
The distribution was fitted by that for an angular
momentum transfer of $\Delta L=2$, calculated by the coupled-channel code 
{\small ECIS97} \cite{ECIS} taking into account the scattering 
angle resolution.
As seen in the figure, the $\Delta L=2$ distribution well reproduced the
experimental one, supporting the $2^+$ assignment for the 1.5~MeV
state.
The {\small ECIS} calculation is almost equivalent to the distorted wave born
approximation, since higher-order processes are negligible in the present
experimental conditions. 
The optical potential parameters were taken from the study of the $^{17}$O +
$^{208}$Pb elastic scattering at 84~MeV/nucleon \cite{Barrette1988182}.
The collective deformation model was employed to obtain a form factor for
nuclear excitation.
The Coulomb- and nuclear-deformation parameters $\beta_C$ and $\beta_N$
were employed to obtain the $B$(E2) value as 
$B({\rm E2})=(3ZeR^2/4\pi )^2 \beta_C^2$.
$\beta_N$ is related to $\beta_C$ by a Bernstein prescription
\cite{Bernstein},
\begin{equation}\label{eq:bNbC}
 \frac{\beta_N}{\beta_C} =
  \frac{1+(b_n^F/b_p^F)(M_n/M_p)}{1+(b_n^F/b_p^F)(N/Z)},
\end{equation}
where $b_{n(p)}^F$ is the interaction strength of a probe $F$ with
neutrons (protons) in the nucleus.
$b_n^F/b_p^F$ is estimated to be 0.81 for the inelastic scattering on Pb
at around 50~MeV/nucleon \cite{Iwasa}.
The $M_n$ was deduced from the adopted $B$(E2) value of the mirror nucleus
$^{28}$Mg \cite{Raman}.
The $B$(E2) value for $^{28}$S was obtained by adjusting $\beta_C$ and 
hence $M_p$ with $\beta_N$ calculated by eq. (\ref{eq:bNbC}) to 
reproduce the experimental angular distribution.
The dashed and dotted curves in Fig. \ref{fig:dsdomega} shows the
Coulomb and nuclear contributions, respectively.
The use of the optical potential determined from the $^{40}$Ar +
$^{208}$Pb scattering \cite{Alamanos198437} gave a 
5.5\% smaller $B$(E2) value.
By taking the average of the results with the two optical potentials,
the $B$(E2;$0^+_{gs} \rightarrow 2^+_1$) value was determined to be 
181(31)~e$^2$fm$^4$.
The associated error included the uncertainty of the measured cross section and
the systematic error due to the choice of optical potentials.
The $B$(E2;$0^+_{gs} \rightarrow 2^+_1$) value for the $^{24}$Mg, 
a contaminant of the secondary beam, was obtained to be 
444(66)~e$^2$fm$^4$ by the same analysis.
This agreed with the adopted value of 432(11)~e$^2$fm$^4$
\cite{Raman}, exhibiting the reliability of the present analysis for $^{28}$S.

%
\begin{figure}[t]
 \includegraphics[clip,keepaspectratio,width=.44\textwidth]{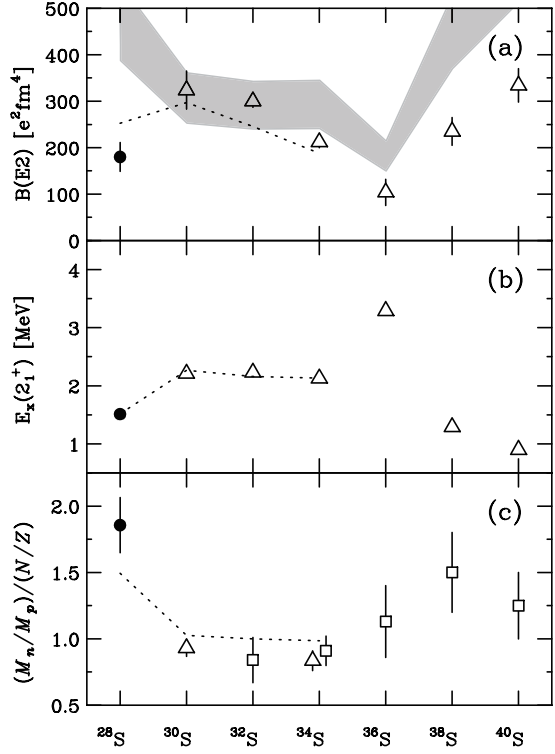}
\caption{\label{fig:Z16N12}
 Plot of the $B$(E2;$0^+_{gs} \rightarrow 2^+_1$) values (a),
 the excitation energies of $2^+_1$ states (b), and
 the double ratio $|M_n/M_p|/(N/Z)$ (c) for sulfur ($Z=16$) isotopes.
 The shell model predictions with the USDB interaction \cite{Brown2} 
 are shown by the dotted curves for each quantity.
 The shaded region represents the $B$(E2) predictions by the empirical
 $B$(E2) systematics \cite{Raman}.
 The present result is represented by the filled circles.
}
\end{figure}
The $B$(E2) and $E_x(2^+_1)$ values for $Z=16$ isotopes are plotted in
Fig. \ref{fig:Z16N12}(a) and (b), respectively.
The filled circles show the present results.
The open triangles for $B$(E2) and $E_x(2^+_1)$ represent known values
for the $Z=16$ isotopes up to $A=40$ \cite{Raman}.
The $B$(E2) value increases from $^{36}$S, the neutron singly-magic
nucleus, to $^{30}$S, and decreases at $^{28}$S.
On the other hand, the $2^+_1$ energy of $^{28}$S is smaller than
those of $^{30-36}$S.
These features contradict the empirical systematics. 
For example, Raman proposed the relation
$B({\rm E2}) = (25.7 \pm 4.5) E_x(2^+_1)^{-1} Z^2 A^{-2/3}$
which is obtained by a global fit to $E_x(2^+_1)$ and $B$(E2) in a wide
range of nuclei \cite{Raman}.
The shaded band in Fig. \ref{fig:Z16N12}(a) represents the $B$(E2) values
calculated by this formula. 
As clearly seen, the present data for $^{28}$S is much smaller than the
expectation of 472(83)~e$^2$fm$^4$.
An explanation of these small $B$(E2) and $E_x(2^+_1)$ is given by the
hindered proton collectivity and the neutron dominance in the 
$0^+_{gs} \rightarrow 2^+_1$ transition. 
A similar mechanism is proposed for $^{16}$C \cite{Elekes,Ong,Wiedeking} and 
$^{136}$Te \cite{Radford, Terasaki}, where small
$B$(E2) and $E_x(2^+_1)$ values in comparison with neighboring isotopes are
observed.

Figure \ref{fig:Z16N12}(c) shows the double ratio $|M_n/M_p|/(N/Z)$ of
the $Z=16$ isotopes.
The filled circle and open triangles show the present result and the
known values, respectively.
They are obtained by the $B$(E2) values of the mirror pairs.
The open squares represent the double ratios obtained by the
combinations of $B$(E2) and the result of $(p,p')$ on the nuclei of
interest \cite{Marechal}.
The ratio for $^{28}$S amounts to 1.9(2) by taking the present result
and adopted $B$(E2) of 350(50)~e$^2$fm$^4$ for the mirror nucleus
$^{28}$Mg \cite{Raman}.
The double ratio of 1.9(2) is significantly larger than unity indicating
again the hindered proton collectivity relative to neutron
and the neutron dominance in the $0^+_{gs} \rightarrow 2^+_1$ transition
in $^{28}$S.
This hindrance can be understood if $^{28}$S is the
proton singly-magic nucleus by the $Z=16$ magicity.
This picture is supported by the larger $B$(E2) value and 
$|M_n/M_p|/(N/Z) \sim 1$ of the neighboring $N=12$ isotones:
356~e$^2$fm$^4$ and 1.05(6) for $^{26}$Si \cite{Raman,Cottle},
and 432(11)~e$^2$fm$^4$ and 0.95(8) for $^{24}$Mg \cite{Raman,Zwieglinski}.
The double ratios of $^{30-36}$S are close to unity, as seen in the
figure, indicating that the hindrance of the proton collectivity does
not appear in these nuclei.
The large double ratios for $^{38, 40}$S can be explained by the neutron skin
effect caused by the $Z=16$ sub-shell closure \cite{Marechal, Alamanos}.

The dotted lines in Fig. \ref{fig:Z16N12} (a)-(c) show shell model
predictions with the USDB effective interaction using the effective charges
of $e_p=1.36$ and $e_n=0.45$ \cite{Brown2, Richter}.
The calculation shows excellent agreement with the experimental 
$E_x(2^+_1)$ values.
The overall tendencies of the $B$(E2) and $|M_n/M_p|/(N/Z)$
are reasonably reproduced.
Especially the sudden decrease of $B$(E2) and increase of
$|M_n/M_p|/(N/Z)$ at $^{28}$S are mostly predicted.
It indicates that the shell model calculation with the USDB interaction
accounts for the phenomena observed in the present study.
It should be note that the model interprets the $N=16$ magicity in
neutron-rich nuclei with the large $s_{1/2}$-$d_{3/2}$ gap, and hence
the $Z=16$ magicity in proton-rich nuclei is inherent in the model
reflecting the isospin symmetry.
Slight difference remaining between the predictions and the experimental
data may require further development of the theory.

In summary, the $B$(E2;$0^+_{gs} \rightarrow 2^+_1$) value for the
proton-rich nucleus $^{28}$S was measured using Coulomb excitation at 
53~MeV/nucleon.
The resultant $B$(E2) value is determined to be 181(31)~e$^2$fm$^4$.
The double ratio $|M_n/M_p|/(N/Z)$ for the $0^+_{gs} \rightarrow 2^+_1$
transition in $^{28}$S is obtained to be 1.9(2), 
by evaluating the $M_n$ value from the known $B$(E2) value of the mirror 
nucleus $^{28}$Mg.
These results show a hindered proton collectivity relative to that of
neutrons in $^{28}$S.
It indicate the emergence of $Z=16$ magicity in the $|T_z|=2$
nucleus $^{28}$S.
The systematics of the $|M_n/M_p|/(N/Z)$ values for the $Z=16$ isotopes
indicates that the hindrance of proton collectivity in proton-rich
region appears only at $^{28}$S.

The authors thank the staff of RIKEN Nishina Center for their work of
the beam operation during the experiment.
One of the authors (Y.T.) is grateful for the support of the Special
Postdoctoral Researcher Program at RIKEN and Research Center for
Measurement in Advanced Science at Rikkyo University.


%

\end{document}